\documentclass[12pt]{article}
\usepackage{graphicx}
\usepackage[utf8]{inputenc}
\usepackage[english]{babel}
\usepackage{amsmath}
\usepackage{amsfonts}
\usepackage{amssymb}

\title{Memory-Induced Transport and Arrest in Flashing Ratchets:
From Superdiffusion to Clustering}

\author{
Karina I. Mazzitello$^{1}$,
Daniel G. Zarlenga$^{2}$,
Constancio M. Arizmendi$^{2}$
}

\date{}

\begin{document}
\maketitle

\begin{center}
{\small
$^{1}$ Laboratorio de Termohidr\'aulica, CNEA, Bariloche, Argentina\\
$^{2}$ Instituto de Investigaciones Cient\'{\i}ficas y Tecnol\'ogicas en Electr\'onica,
Universidad Nacional de Mar del Plata, Argentina
}
\end{center}

\begin{abstract}
We investigate the transport properties of particles driven by colored noise in a flashing ratchet potential, focusing on both non-interacting and single-file interacting regimes. The model incorporates memory effects via a non-Markovian friction kernel, leading to superdiffusive dynamics and enhanced currents in the absence of interactions. However, when particles are constrained to single-file motion with hard-core repulsion, the same non-Markovian noise induces a dynamical transition: initial superdiffusion gives way to the formation of static clusters, ultimately suppressing net current. This transition occurs without a critical density and results from the interplay between noise persistence and the ratchet’s potential. Our numerical results reveal a universal scaling behavior for the mean square displacement across densities, suggesting robustness of the clustering mechanism. These findings have potential implications for transport in crowded or confined systems such as colloidal suspensions, molecular motors in cellular environments, or microfluidic devices, where controlling noise and crowding can be used to tune transport efficiency.
\end{abstract}

\section{Introduction}
The interplay of noise, spatial asymmetry, and interparticle interactions represents a fundamental aspect in the study of non-equilibrium statistical physics, with significant implications across biological systems, soft condensed matter, and nanoscale devices. In particular, flashing ratchet models--where an asymmetric potential is periodically switched on and off--have emerged as fundamental frameworks to understand the rectification of stochastic fluctuations into directed transport without net external force, a phenomenon often described as Brownian motor action \cite{Reimann2002, Astumian2002}.\\

The role of noise in these systems is especially pivotal. While much of classical stochastic transport theory has focused on white (delta-correlated) noise, real-world systems frequently involve temporally correlated fluctuations--commonly referred to as colored noise--which introduce memory into the dynamics \cite{van_Kampen1992, Hanggi2009, Jung1987}. The inclusion of colored noise in flashing ratchets has been shown to produce markedly different behaviors, such as superdiffusive transport regimes and significant enhancement of net current due to persistent stochastic driving \cite{Bao2003, Bao2004, Goychuk2012,Sokolov2012}. These features align with theoretical predictions from generalized Langevin models incorporating non-Markovian friction kernels, which better reflect the viscoelastic and correlated environments often encountered in biological and soft-matter contexts \cite{Kupferman2004,Siegle2010, Kou2004}.\\

While extensive studies have characterized the behavior of non-interacting particles under colored noise in ratchet systems \cite{Lindner2004}, less is understood about how strong interactions—particularly those arising from spatial confinement--affect the collective dynamics. In one-dimensional (1D) geometries with hard-core repulsion, particles are restricted to single-file motion, where overtaking is prohibited. This constraint alters diffusion fundamentally: the mean square displacement (MSD) scales sublinearly with time, and anomalous clustering phenomena can emerge \cite{Percus1974, Harris1965, Levitt1973}.\\

To investigate these effects, we study the dynamics of interacting particles within a flashing ratchet potential driven by colored noise, modeled using a generalized Langevin equation with a bi-exponential memory kernel. This setup emulates realistic environments where crowding and temporal correlations coexist, such as intracellular transport along cytoskeletal filaments, nanopore diffusion, and colloidal particles in microfluidic channels \cite{Reguera2006, Marbach2015,Ai2005,Burada2009}.

We focus on how noise persistence and geometric constraints collectively influence transport in dilute  systems. In the absence of interactions, colored noise leads to ballistic or superdiffusive behavior, boosting net transport. However, when particles are confined to a single-file regime, we uncover a striking dynamical transition: long-time motion is suppressed due to cluster formation. Remarkably, this transition does not require a critical particle density and appears robust across parameter regimes.\\

These findings provide fresh insights into how non-Markovian dynamics interact with spatial constraints to generate nontrivial collective behavior. They may inform future experimental and theoretical work in designing efficient transport systems, controlling diffusion in crowded environments, and understanding naturally occurring transport phenomena in biological and synthetic systems \cite{Eshuis2005,Benichou2011,Malgaretti2013,Lutz2004}.\\

\section{MODEL}
In this model, particles of identical mass $m$ are subjected to a flashing ratchet and a colored noise source, as described in \cite{Bao2003}. The motion of each particle is governed by the following Langevin equation:
\begin{equation} \label{langevin_equation}
	m\: \ddot{x}(t)+ \int_{0}^{t}\eta(t-t') \: \dot{x}(t') \: dt' = \varepsilon(t) + f(x,t)
\end{equation}
where the left-hand side represents the particle's inertia and a frictional force with memory effects, while the right-hand side consists of a fluctuating noise term $\epsilon(t)$ and the flashing ratchet force $f(x,t)$. The memory kernel $\eta(t-t')$, which characterizes the frictional force, is defined as:
\begin{equation} \label{memory_kernel}
	\eta(t-t')=A \left[ \frac{1}{\tau_{2}}\exp\left(-\frac{\left| t-t'\right| }{\tau_{2}} \right)-\frac{1}{\tau_{1}}\exp\left(-\frac{\left| t-t'\right| }{\tau_{1}} \right)  \right]\mbox{,} 
\end{equation}
where  $\tau_{1}$ and $\tau_{2}$ are time constants with $\tau_{1}>\tau_{2}$. The prefactor $A$ depends on the friction coefficient $\eta_0$ and is given by:
\begin{equation} \label{A_constant}
	A=\eta_{0}\frac{\tau_{1}^{2}}{\tau_{1}^{2}-\tau_{2}^{2}}\mbox{.}
\end{equation}
The memory kernel $\eta(t-t')$ indicates that the particle’s current motion is influenced by its past states, which leads to non-Markovian dynamics. This temporal correlation is crucial in producing superdiffusive behavior, where the particle’s displacement square mean grows faster than it would under normal diffusion.\\
The noise term $\epsilon(t)$ represents thermal fluctuations due to the interaction between the particle and its environment. These fluctuations satisfy the fluctuation-dissipation relation:
\begin{equation}\label{thermal_noise}
	\left\langle \varepsilon(t) \varepsilon(t') \right\rangle = k_B \: T \: \eta(t-t')\mbox{,}
\end{equation}
where $k_B$ is the Boltzmann constant, $T$ is the absolute temperature, and $\langle...\rangle$ denotes the ensemble average. This relationship reflects that friction and thermal noise share a common physical origin, the interaction of the particle with its surroundings. Besides, we assume that $\eta(t)$ is a zero-mean stationary process.\\
The flashing ratchet force $f(x,t)$ in Eq.~(\ref{langevin_equation}) is defined by:
\begin{equation} \label{force}
	f(x,t)=-\frac{dU(x)}{dx}  F(t)\mbox{,}
\end{equation}
where $U(x)$ is a spatially periodic sawtooth potential with period $L$, given by:
\begin{equation} \label{ratchet_force_equation}
U(x) =
\left\lbrace  {\begin{matrix}
		-\frac{U_0}{(1-\alpha)L}x & & 0<x<(1-\alpha)L \\
		\\
		\frac{U_0}{\alpha L}x & & (1-\alpha)L<x<L
\end{matrix} }  \right.
\end{equation}
with $U_0$ being the potential barrier height and $\alpha$ describing the asymmetry of the potential. The ratchet alternates between "on" and "off" states through the time-dependent function $F(t)$:
\begin{equation} \label{ratchet_onoff_equation}
    F(t) =
    \left\lbrace
    \begin{matrix}
        1 & \text{if } 0 < t < t_p / 2 \\ 
        0 & \text{if } t_p / 2 < t < t_p
    \end{matrix}
    \right. , 
\end{equation}
where $t_p$ is the period of the ratchet switching cycle.\\
For simplicity and without loss of generality, we set the drag coefficient $\eta_0=1$, and measure energy and length in units of $k_B$ and $L$ respectively, meaning $k_B=1$ and $L=1$.\\
The noise described by Eq.~(\ref{memory_kernel}) contains two additive components corresponding to the memory times $\tau_1$ and $\tau_2$, leading to a power spectral density:
\begin{equation} \label{memory_kernel_power_spectrum}
	S(\omega)=k_B \: T \: \frac{2 \: \eta_{0} \: \tau_{1}^{2} \: \omega^{2}}{\left(1+\tau_{1}^{2} \: \omega^{2} \right) \left(1+\tau_{2}^{2} \: \omega^{2} \right)}
\end{equation}
This indicates that the noise spectrum has two Lorentzian peaks centered around the characteristic frequencies $\omega_1=1/\tau_1$ and $\omega_2=1/\tau_2$. Particles oscillating near these frequencies experience enhanced motion, which promotes superdiffusion as we will see in the next section.
We solve Eq.~(\ref{langevin_equation}) numerically following the approach detailed in \cite{Bao2004}.\\ Initially, particles are uniformly distributed under periodic boundary conditions in a system consisting of multiple sawtooth-shaped potentials. Throughout the study, and without loss of generality, we set the particle mass to $m = 1$, the ratchet potential amplitude to $U_0 = 1$, the flashing period to $t_p=0.6$, the temperature to $T = 0.001$ and the bi-exponential memory times to $\tau_1=0.5$ and $\tau_2 = 0.05$. We compare systems driven by bi-exponential memory kernel with those driven by a single memory kernel, under a configuration where single-file constraints apply only to interacting particles, while non-interacting particles propagate freely. In a single-file regime, particles interact via hard-core repulsion, which prevents them from bypassing each other during attempted overtakes.\\
We computed the mean square displacement ($MSD$) as
\begin{equation}
MSD=\langle \left(x(t)-x_0\right)^2\rangle \;\,
\end{equation}
where $\langle ...\rangle $ denotes the average taken over all particles, and $\left(x(t)-x_0\right)$ represents the displacement of a particle from its initial position.  Additionally, we calculate the particle current $J$ defined as
\begin{equation}
 J=\rho \frac{\langle x(t + \Delta t)-x(t)\rangle }{\Delta t}\;\mbox{,}
\end{equation}
with $\rho$ the particle density. We also evaluated the mean absolute displacement, $\langle |(x(t)-x_0)|\rangle $.\\
\section{RESULTS}
\subsection{Superdiffusive Motion of Non-Interacting Particles}
\begin{figure}[htbp]
\centering
\includegraphics[width=0.49\textwidth]{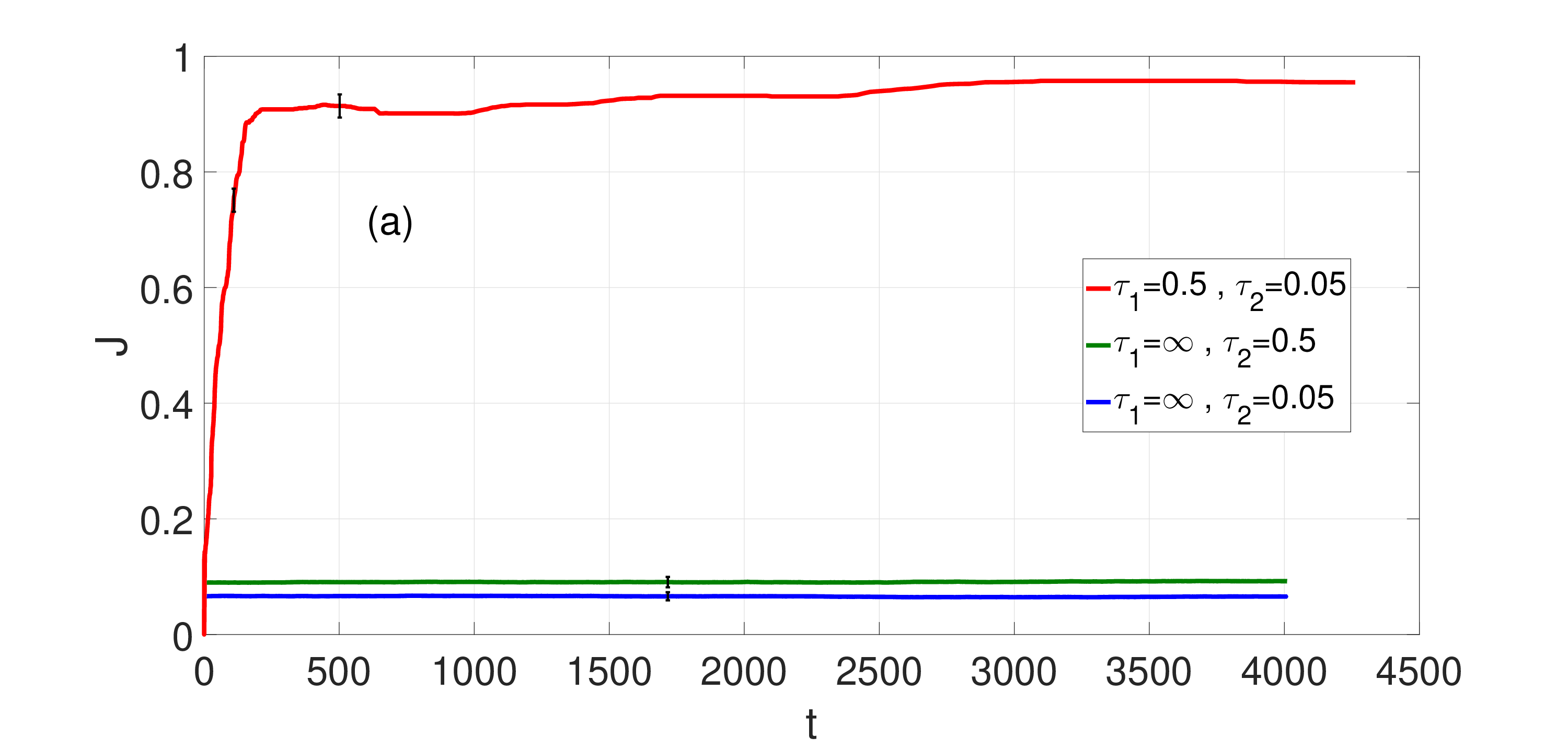}
\hfill
\includegraphics[width=0.49\textwidth]{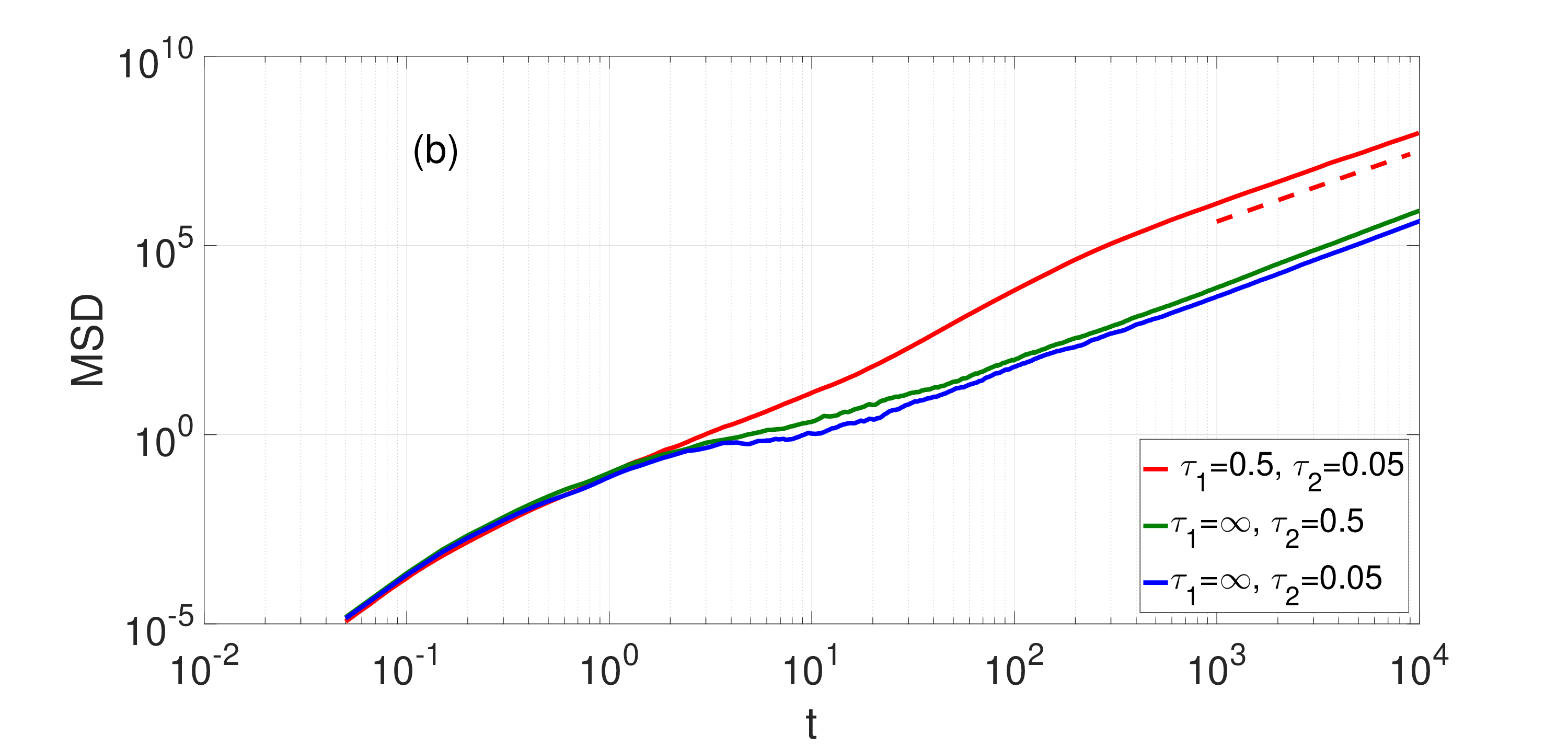}
\caption{\label{fig1}
Comparison of (a) current
$J$ and (b) mean-squared displacement ($MSD$) as functions of time for non-interacting particles subjected to colored noise with two characteristic frequencies (red line) and with a single characteristic frequency (green and blue lines). In all cases, the particle density is fixed at
$\rho = 1$. The two-frequency noise yields higher transport efficiency. The dashed line in panel (b) has slope 1.9 and is shown as a reference. (Representative error bars are shown in panel (a)).}
\end{figure}

The model described above has been extensively investigated in previous studies for non-interacting particles~\cite{Bao2003, Bao2004, Bao1999}. The colored noise introduced in Eq.~(\ref{memory_kernel}) is characterized by two distinct frequencies. Suppressing one of these frequencies, achieved by taking the limit $\tau_1 \to \infty$ while keeping $\tau_2$ finite, leads to a pronounced reduction in particle transport through the ratchet potential. This effect is illustrated in Fig.~\ref{fig1}(a): the green and blue curves show the current $J$ as a function of time for a single frequency scenario corresponding to a memory time $\tau_2 = 0.5$ and $\tau_2 = 0.05$, respectively. In contrast, the red curve correponds to the two-frequency case with memory times $\tau_1 = 0.5$ and $\tau_2 = 0.05$, all obtained for non-interacting particles.\\
Both types of colored noise, bi-exponential (two kernels) and single-exponential (one kernel), induce superdiffusive motion in non-interacting particles, as shown in Fig.~\ref{fig1}(b). The plot displays the mean-square displacement ($MSD$) as a function of time on a log–log scale for the same memory times considered in panel (a). In the steady state, the superdiffusive regimes are characterized by slopes approaching 2.0. This anomalous diffusion arises from the temporal nonlocality introduced by the memory kernel, which contributes non-Markovian corrections to the particle velocity and thereby sustains superdiffusive dynamics.\\
The single-exponential kernel has been extensively studied in the context of rocking ratchets, where the memory time effectively reduces the mass and thereby increases the current as the memory time grows \cite{Spiechowicz2025}. Accordingly, $J$ for $\tau_2 = 0.5$ (green curve in Fig.~1(a)) exceeds $J$ for $\tau_2 = 0.05$ (blue curve). In comparison, the bi-exponential kernel drives the particles at two distinct frequencies, which further enhances their ability to overcome potential barriers. Here, results are shown for a fixed $t_p = 0.6$, but this enhancement persists across the full range of flashing ratchet frequencies, approaching at very high or very low values ($0.2 \leq t_p \leq 3.6$).\\
\subsection{Motion of interacting Particles}
\begin{figure}[htbp]
\centering
\includegraphics[width=0.49\textwidth]{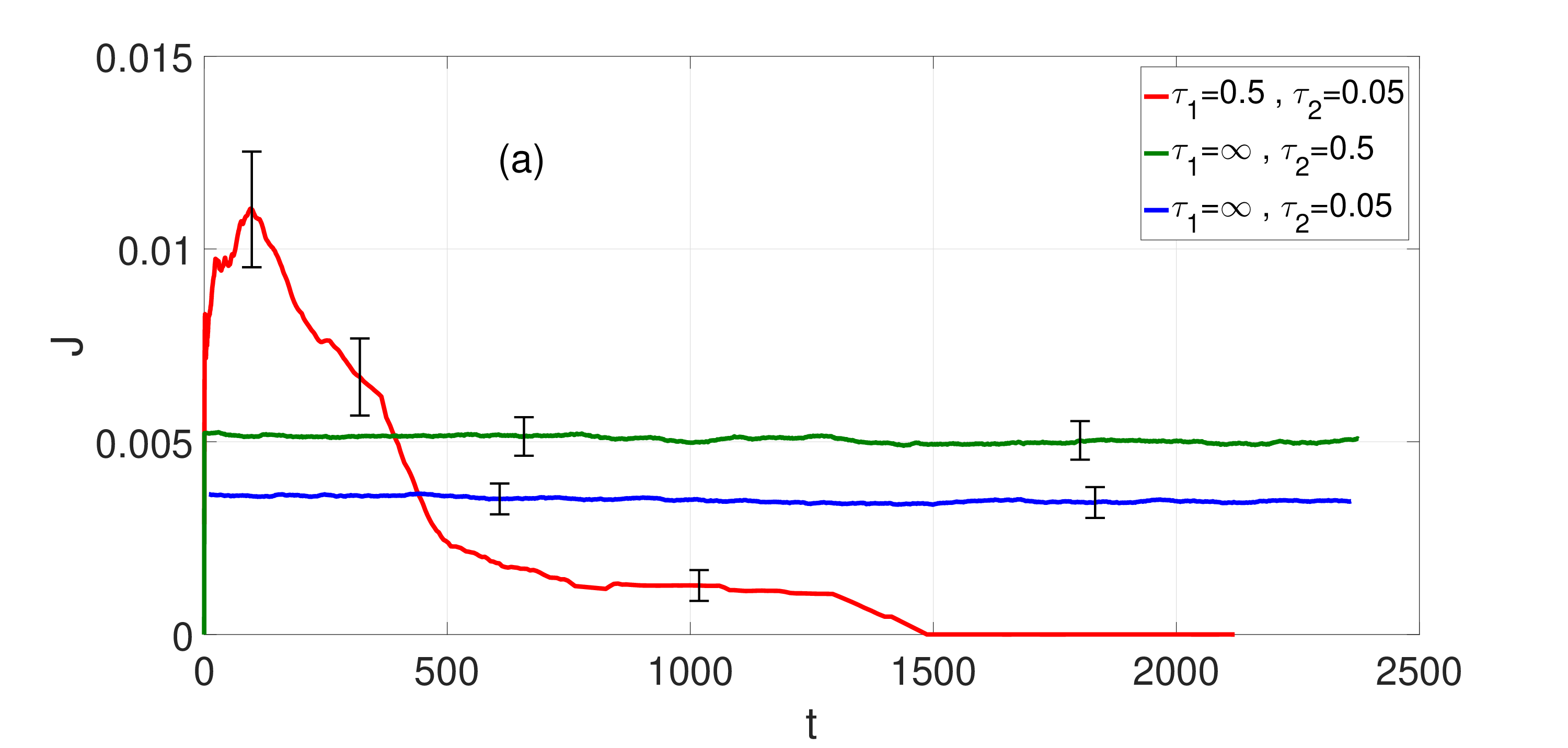}
\hfill
\includegraphics[width=0.49\textwidth]{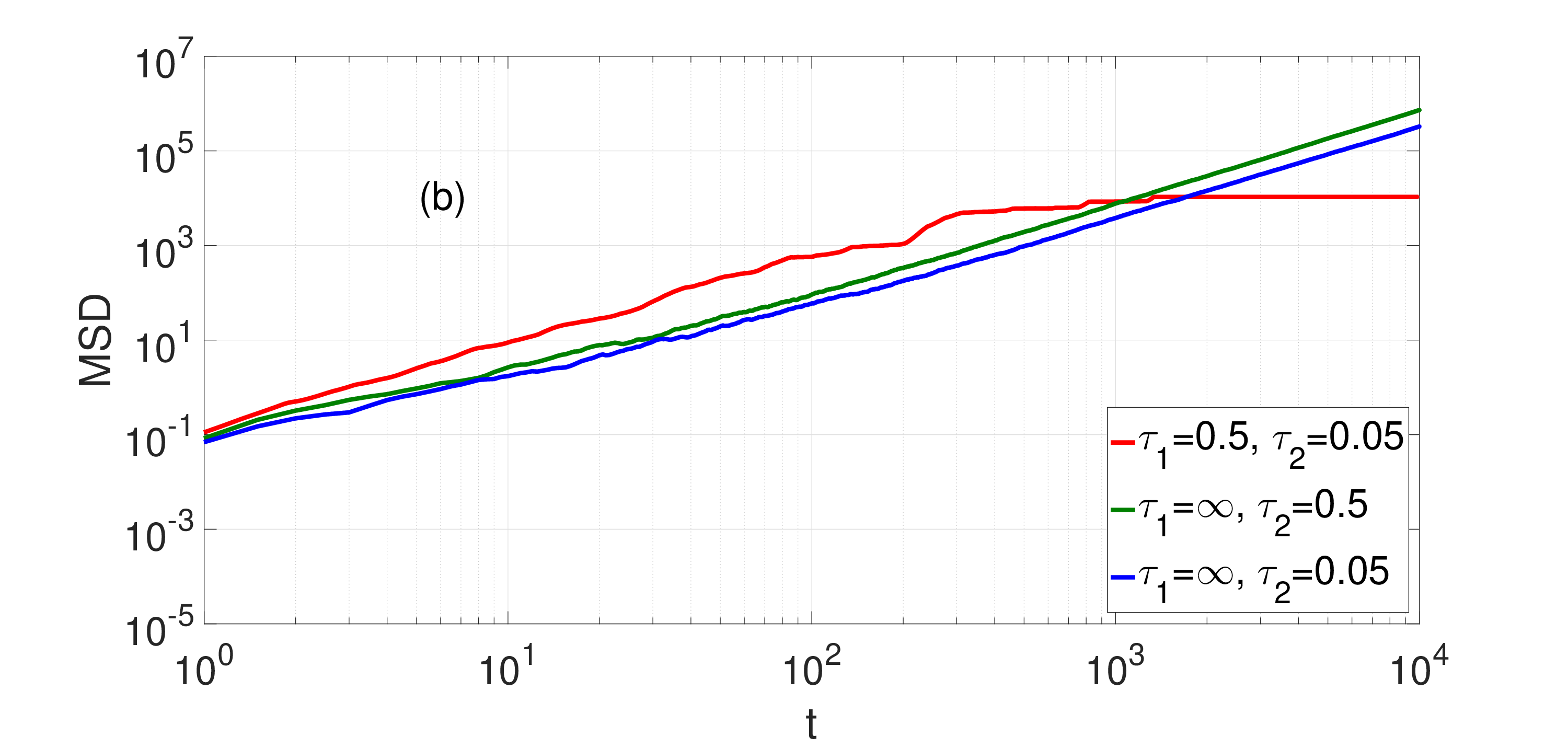}
\caption{\label{fig2} Comparison of (a) $J$ and (b) the $MSD$ vs. time for interacting particles under colored noise with two characteristic frequencies (red line) and with one characteristic frequency (green and blue lines). The particle density $\rho = 0.06$ for all cases. In the first case, particle interactions suppress $J$, resulting in a plateau of the $MSD$. (Representative error bars are shown in panel (a).}
\end{figure}

The dynamics change markedly when particle interactions are included, specifically through hardcore repulsion. Fig.~\ref{fig2} compares single-file systems driven by colored noise with two characteristic frequencies (red curves) to those driven by a single characteristic frequency (blue and green curves) in a flashing ratchet setup. Panels (a) and (b) show the current $J$ and the mean square displacement ($MSD$), respectively, as functions of time for a dilute system with $\rho = 0.06$.\\
At short times, the probability of particle interactions is low, and the current under two-frequency colored noise (red curve in Fig.~\ref{fig2}(a)) is substantially greater than in the single-frequency cases (blue curve in Fig.~\ref{fig2}(a)). However, once the particles interact with each other, the current for the two-frequency case drops dramatically to zero. Thus, the same colored noise that produces high currents in the non-interacting system completely suppresses net motion in single-file configurations (see previous section).\\
This suppression is also reflected in the $MSD$: under two-frequency noise, the $MSD$ reaches a plateau, indicating confinement of particle motion, whereas under single-frequency noise, particles continue to diffuse over time (Fig.~\ref{fig2}(b)).\\
\subsection{Cluster formation in interacting single-file systems}
\begin{figure}[htp]
\centering
\includegraphics[width=21.pc]{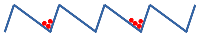}
\caption{\label{Esquema} Schematic illustration of cluster formation in a single-file system of particles subjected to colored noise with two characteristic frequencies.}
\end{figure}

\begin{figure}[htp]
\centering
\includegraphics[width=21.pc]{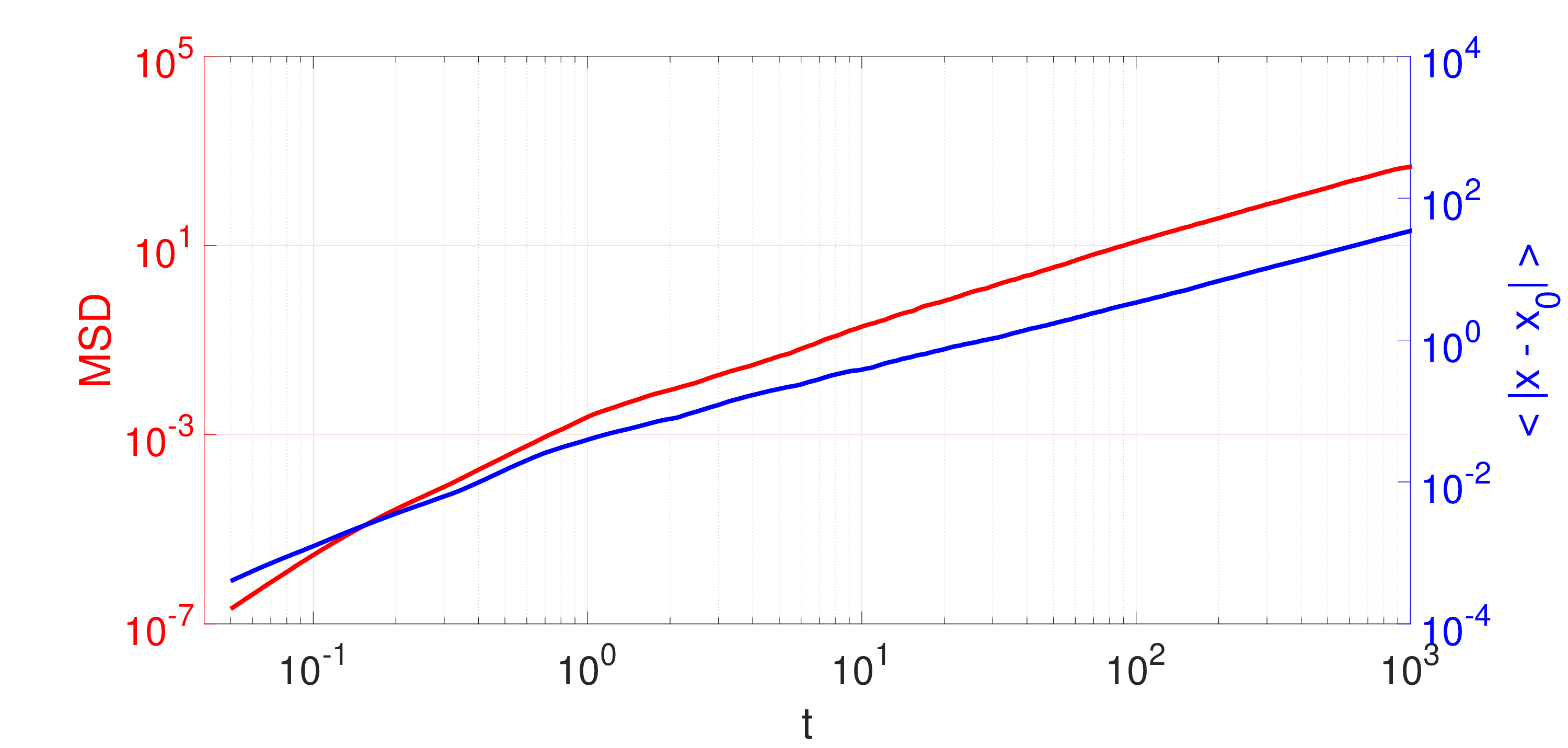}
\caption{\label{fig4} MSD and $\langle |(x-x_0)|\rangle $ versus time for $U_0 = 0$ and colored noise with two characteristic frequencies on a log–log scale, for a single-file system (red line) and free particles (blue line). The particle density $\rho = 0.06$ for both cases. The behavior with time predicted by Eq.~(\ref{single-file}) is satisfied in the steady state.}
\end{figure}

A remarkable property of many single-file systems is that their steady-state dynamics can be described entirely in terms of the motion of individual non-interacting particles. Hahn~\cite{Hahn1995} showed that the $MSD$ of interacting particles in such systems can be expressed as the product of the mean free path of interacting particles, $\lambda_{path}$, and the mean displacement of free (non-interacting) particles, $\langle |(x-x_0)|\rangle $:\\
\begin{equation}\label{single-file}
MSD = \lambda_{path} \langle |(x-x_0)|\rangle \;\mbox{.}
\end{equation}
Simulation results indicate that Eq.~(\ref{single-file}) is valid for single-frequency noise, but fails in the two-frequency case, where particles become locally confined and form clusters at the minima of the ratchet potential (see Fig.\ref{Esquema}).\\
When the ratchet potential is switched off,
Eq.~(\ref{single-file}) is recovered, and the system exhibits
standard single-file diffusion. This behavior is shown in Fig.~\ref{fig4},
where the $MSD$ (solid line) for interacting particles and the average
absolute displacement $\langle |(x-x_0)|\rangle $ (dashed line) for free
particles are compared on a log–log scale. In this case, with particles subject only to two-frequency colored noise, no clustering occurs. This demonstrates that both the ratchet potential and the persistence
associated with this type of noise are required for cluster formation.\\
As illustrated schematically in Fig.~\ref{Esquema},
clusters typically involve more than two particles. Simulations
show that while two particles occupying the same potential well
can still diffuse, pairs of particles separated by three ratchet
periods tend to form stable clusters at long times. This clustering
leads to a vanishing current $J$ and causes the $MSD$ to saturate,
reaching a plateau.\\
\begin{figure}[htbp]
\centering
\includegraphics[width=0.49\textwidth]{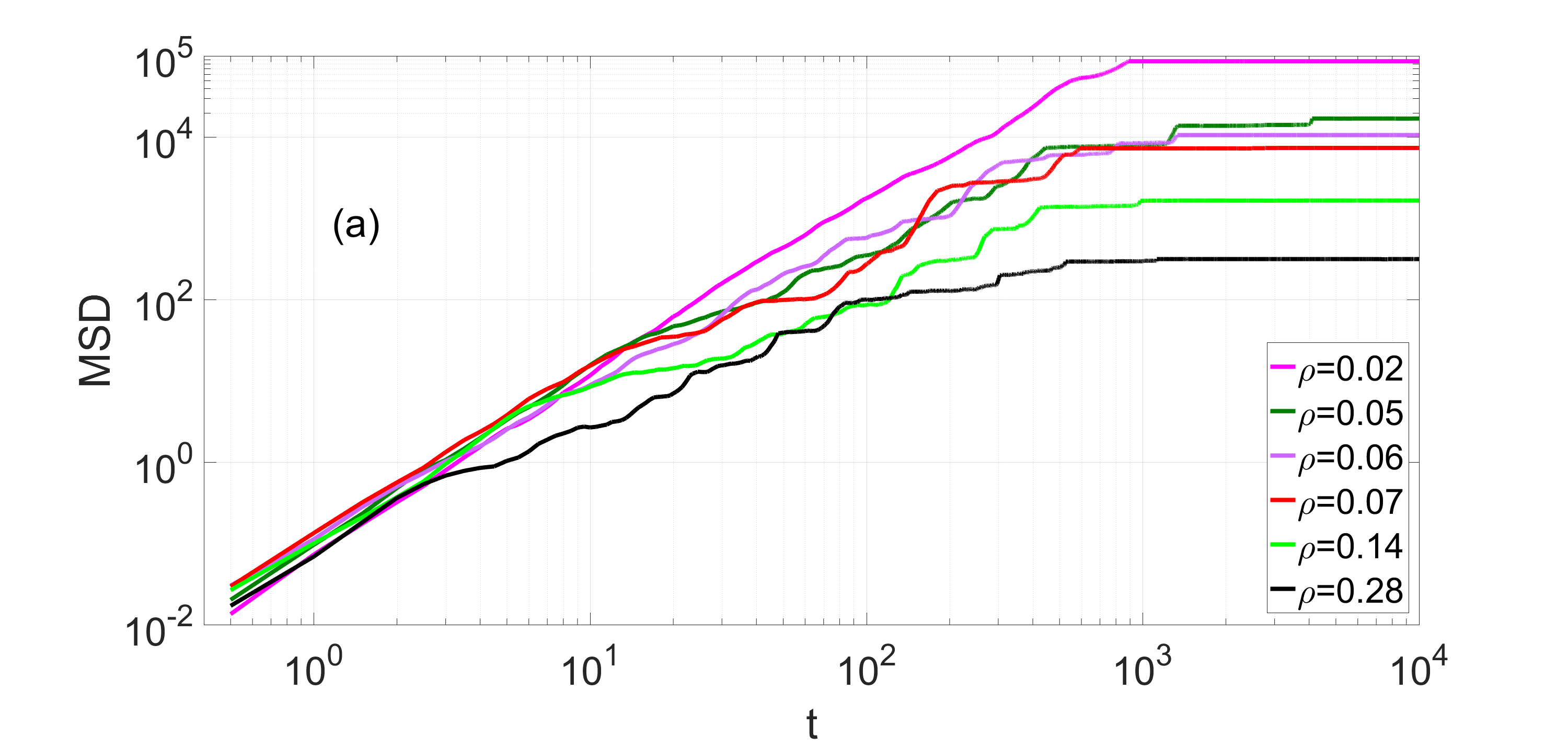}
\hfill
\includegraphics[width=0.49\textwidth]{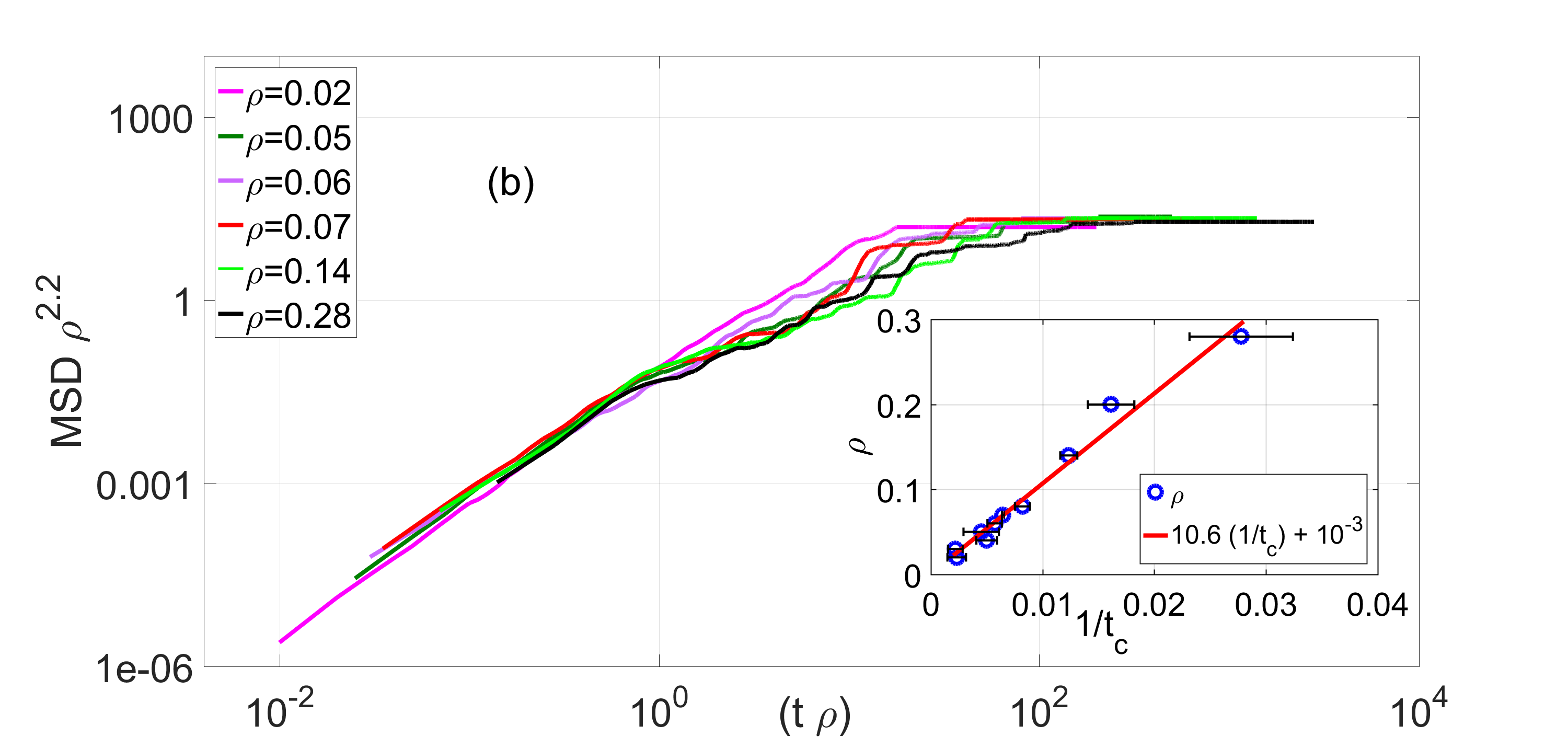}
\caption{\label{fig5} (a) $MSD$ versus time for $t_p/2 = 0.3$, $\alpha = 0.9$, and different particle densities $\rho$ under two-frequency colored noise. (b) Scaled $MSD$ as a function of $t\rho$. Inset: $\rho$ vs. $1/t_c$ with linear fit (red line).}
\end{figure}
Figure~\ref{fig5}(a) shows the $MSD$ as a function of time for single-file systems with different particle densities under two-frequency colored noise. In all cases, the MSD reaches a maximum value that depends on particle density. Each curve corresponds to the average over one sample including all particles in the system; before reaching the plateau, the curves display small steps, foreshadowing the eventual immobilization of the particles. The crossover time $t_c$, which separates the superdiffusive and clustering behaviors, is found to be inversely proportional to the density. A scaling analysis, presented in Fig.~\ref{fig5}(b), captures the behavior at both short and long times and confirms the absence of a finite critical density for cluster formation. The inset of Fig.~\ref{fig5}(b) plots $\rho$ versus $1/t_c$; the linear fit yields an intercept close to zero, indicating that a finite crossover time exists for all particle densities.\\
Additional simulations were performed by varying the ratchet potential asymmetry $\alpha$ ($0.6 \leq \alpha \leq 0.9$) and the flashing period $t_p$ ($0.2 \leq t_p \leq 3.6$), confirming the robustness of these results. In this flashing ratchet model driven by two-frequency colored noise (Eq.~(\ref{memory_kernel}), the current vanishes and the $MSD$ saturates at long times, demonstrating that clustering emerges as a generic outcome of the dynamics.\\
\section{Conclusions}
We investigated the transport properties of particles subjected to a flashing ratchet potential and driven by colored noise with memory effects. Our study focuses on a one-dimensional setup comprising both single-file interacting particles, where overtaking is forbidden, and non-interacting particles. Using a generalized Langevin equation with a bi-exponential memory kernel, we numerically explored the emergence of superdiffusive motion and its impact on current generation.\\
For non-interacting particles, the interplay between the colored noise and the ratchet potential results in superdiffusion and enhanced currents. Compared to systems driven by single exponential memory kernel, particles subjected to colored noise with a bi-exponential memory kernel exhibit markedly higher transport efficiency. This improvement arises because the bi-exponential kernel excites the particles at two distinct frequencies, thereby increasing their ability to overcome potential barriers relative to the single-frequency case.\\
However, when interactions are introduced through hard-core repulsion, the dynamics change dramatically. Initially superdiffusive, the motion of particles becomes constrained by the single-file configuration. Over time, the system exhibits clustering behavior, where groups of particles become trapped in the minima of the ratchet potential. This collective trapping leads to the eventual suppression of the particle current, driving the system to a dynamically arrested state.\\
We showed that this clustering phenomenon is specific to the combination of colored noise and the ratchet potential, as neither component alone leads to such behavior. Our results further demonstrate that the transition to this arrested state occurs across a range of particle densities, with no evidence of a critical threshold.\\
In summary, our work reveals a rich dynamical interplay between memory-induced superdiffusion and interaction-driven confinement. While colored noise with a bi-exponential memory kernel enhances transport in non-interacting systems in flashing ratchet, it paradoxically inhibits long-time mobility in single-file environments by promoting the formation of stable clusters. These findings highlight the complex and sometimes counterintuitive consequences of non-Markovian noise in systems with constrained geometries, and may be relevant to understanding transport in biological channels, porous media, or synthetic nanodevices with ratchet-like features.\\
\bibliographystyle{unsrt}
\bibliography{DGZbibratchet2}
\end{document}